\documentclass[smallcondensed]{svjour3}

\smartqed
\usepackage{graphicx}

\begin{document}

\title{Wormhole Geometries In $f(R,T)$ Gravity}

\author{Tahereh Azizi}
\institute{Department of Physics, Faculty of Basic Sciences,\\
University of Mazandaran, P. O. Box 47416-95447, Babolsar, IRAN
\\\email{t.azizi@umz.ac.ir}}
\maketitle

\begin{abstract}
We study wormhole solutions in the framework of $f(R,T)$ gravity
where $R$ is the scalar curvature, and $T$ is the trace of the
stress-energy tensor of the matter. We have obtained the shape
function of the wormhole by specifying an equation of state for the
matter field and imposing the flaring out condition at the throat.
We show that in this modified gravity scenario, the matter threading
the wormhole may satisfy the energy conditions, so it is the
effective stress-energy that is responsible for violation of the null energy condition.\\


\end{abstract}


\section{Introduction}
One of the great discoveries in modern cosmology has been the
accelerating expansion of the universe \cite{Rie98,Per99,Spe03}.
Among the various interesting possibilities invoked in order to
explain the cosmic speed up, $f(R)$ modified gravity models ($R$ is
the scalar curvature) have attracted a lot of attention(see
\cite{Cap03,Noj03a,Noj03b,Car04,Noj06,Mul06,All06,Sta07,Noj07,Noz09,Sot10,Noj11}
and references therein). The Einstein field equations of General
Relativity were firstly derived from an action principle by Hilbert,
by adopting a linear function of the scalar curvature, $R$, in the
gravitational Lagrangian density. An extension of the
Einstein-Hilbert Lagrangian to the $f(R)$ gravity scenario can be
performed naturally as there is no a priori reason why the
gravitational action should be linear in the Ricci scalar $R$.
Furthermore, higher order terms can naturally appear in low energy
effective Lagrangians of quantum gravity and string theory.

In Ref. \cite{Har11}, a generalization of $f(R)$ modified theories
of gravity was proposed by including in the theory a coupling of an
arbitrary function of the Ricci scalar with the trace of the
stress-energy tensor $T$ , i.e. $f(R,T)$ gravity. They have
investigated some astrophysical and cosmological aspects of the
scenario by choosing several functional forms for $f$ (see also
\cite{Jam12,Hou12}).

In this paper, we explore the possibility whether static and
spherically symmetric traversable wormhole geometries are supported
by $f(R,T)$ gravity. We show that, in this modified theory, static
wormhole throats respecting the null energy condition (NEC) can
exist. Note that as is widely known, traversable wormholes as
solutions to the Einstein equations can only exist with exotic
matter which violates the null energy condition
\cite{Ell73,Bro73,Cle81,Mor88,Vis95,Hoc98}. The null energy
condition holds if $T_{\mu\nu}n^\mu n^\nu \ge 0$ for any null vector
field $n^\mu$. The search of realistic physical models providing the
wormhole existence represents an important direction in wormhole
physics. Various models of such kind include scalar fields
\cite{Bar99,Kas08}; wormholes geometries induced by quantum
effects\cite{Noj99a,Noj99b}, wormhole solutions in semi-classical
gravity \cite{Sus10,Gar07}; solutions in modified gravity
\cite{Lob09,DeB11,Gar10,Gar11}; wormholes on the brane
\cite{Anc00,Bro03}; wormholes supported by generalized Chaplygin gas
\cite{Lob06}; wormhole solutions in Einstein-Gauss-Bonnet theory
\cite{Ric07,Kan11}; modified teleparallel gravity \cite{Boe12}, etc
(for instance see \cite{Lem03} and references therein).

This paper is organized as follows. In section II, we present a
brief review of the fundamental concepts of $f(R,T)$ gravity, the
action of the scenario and equations of motion. We explore the
wormhole geometries in $f(R,T)$ gravity in section III. Firstly, we
introduce the space-time metric and the necessary conditions to have
a traversable wormhole solution. In the second stage, we choose a
special functional form for $f$ and investigate the solutions of the
gravitational field equations. By specifying an equation of state
for the matter field, we obtain the shape function  of the wormhole.
We impose that the matter threading the wormhole satisfies the
energy conditions. So, it is the effective stress-energy tensor that
is responsible for the violation of the null energy condition.
Finally, our summery and conclusions are presented in section IV.

\section{$f(R,T)$ gravity}
The action of $f(R,T)$ gravity is of the following form \cite{Har11}
\begin{equation}
{\cal{S}}=\frac{1}{16\pi G}\int d^{4}x\sqrt{-g}\,
f\left(R,T\right)+\int
d^{4}x\sqrt{-g}\,{{\cal{L}}_{m}}\label{action}
\end{equation}
Here $f\left(R,T\right)$ is an arbitrary function of the scalar
curvature, $R=R_{\mu}^{\mu}$, and the trace $T=T_{\mu}^{\mu}$ of the
stress-energy tensor of the matter, $T_{\mu \nu}$. ${\cal{L}}_{m}$
is the Lagrangian density of the matter and is related to the
stress-energy tensor as follows
\begin{equation}
T_{\mu \nu }=-\frac{2}{\sqrt{-g}} \frac{\delta \left(
\sqrt{-g}{\cal{L}}_{m}\right) }{\delta g^{\mu \nu }}\, .
\end{equation}
Assuming that the Lagrangian density of matter ${\cal{L}}_{m}$
depends only on the metric $g_{\mu \nu }$, we deduce that
\begin{equation}
T_{\mu \nu }=g_{\mu \nu }{\cal{L}}_{m}-2 \frac{\partial
{\cal{L}}_{m}}{\partial g^{\mu\nu}}\, .
\end{equation}
Varying the action (\ref{action}) with respect to the metric
provides the field equations of $f(R,T)$ gravity \cite{Har11}
$$
f_{R}\left(R,T\right) \left( R_{\mu \nu }-\frac{1}{3}Rg_{\mu \nu
}\right) +\frac{1}{6} f\left( R,T\right)  g_{\mu\nu } =$$
$$8\pi G\left(T_{\mu \nu}-\frac{1}{3}T g_{\mu \nu}\right)-f_{T}\left(
R,T\right) \left( T_{\mu \nu }-\frac{1}{3}T g_{\mu \nu
}\right)\nonumber
$$
\begin{equation}
-f_T\left(R,T\right)\left(\Theta_{\mu \nu}-\frac{1}{3} \Theta
g_{\mu\nu}\right) + \nabla _{\mu }\nabla _{\nu
}f_{R}\left(R,T\right)\,\label{field:eq1} .
\end{equation}
where we have denoted  $f_{R}\left( R,T\right) =\partial
f\left(R,T\right) /\partial R$ and $f_T\left( R,T\right) =\partial
f\left(R,T\right) /\partial T$, respectively and
\begin{equation}
\Theta_{\mu \nu}\equiv g^{\alpha \beta }\frac{\delta T_{\alpha \beta
}}{\delta g^{\mu \nu}}\,\label{tetta} .
\end{equation}

 In this paper, we assume that the matter Lagrangian is given by
${\cal{L}}_{m}=-\rho$, where $\rho $ is the energy density (see
\cite{Bro93,Far99,Gar10}). As a result, Equ. (\ref{tetta}) takes the
following form
\begin{equation}
\Theta_{\mu \nu }=-2T_{\mu \nu }-\rho g_{\mu \nu }\, .\label{tetta2}
\end{equation}

\section{Whormhole geometries in $f(R,T)$ gravity}

\subsection{ Spacetime metric and the Gravitational Field Equations}

For the wormhole metric, we consider the following line element
\cite{Mor88}
\begin{equation}
ds^2=-e^{2\Phi(r)}dt^2+\frac{dr^2}{1-b(r)/r}+r^2\,(d\theta^2 +\sin
^2{\theta} \, d\phi ^2) \,,
\end{equation}
where $\Phi(r)$ and $b(r)$ are two arbitrary functions of $r$ known
as the redshift and shape functions respectively. The radial
coordinate $r$ is non-monotonic such that it decreases from infinity
to a minimum value $r_0$, representing the location of the throat of
the wormhole, where $b(r_0)=r_0$, and then it increases from $r_0$
towards infinity. To have a traversable wormhole solution, it is
necessary to impose the flaring out condition, given by
$(b-b^{\prime}r)/b^{2}>0$, \cite{Mor88,Lob09}, and at the throat
with $b(r_{0})=r=r_{0}$, the conditions $b^{\prime}(r_{0})<1$ and
$1-b(r)/r>0$ are imposed. For the wormhole to be traversable, one
must demand that there are no horizons present, which are identified
as the surfaces with $e^{2\Phi}\rightarrow0$, so that $\Phi(r)$ must
be finite everywhere. In the following analysis, for simplicity, we
consider that the redshift function is constant so, $\Phi'=0$.

 In the rest of this paper, we assume that
$f\left(R,T\right)=R+2f(T)$, where $f(T)$ is an arbitrary function
of the trace of the stress-energy tensor. The gravitational field
equations (\ref{field:eq1}), by the definition (\ref{tetta2}) take
the following form
\begin{equation}
R_{\mu\nu}-\frac{1}{2}Rg_{\mu\nu}=8\pi G T_{\mu\nu}
+2FT_{\mu\nu}+\left(2\rho F+f\right)g_{\mu \nu }\,.
\end{equation}
where $f=f(T)$ and $F=\frac{df}{dT}$. Supposing $8\pi G\equiv1$,
this equation can be recast in the form
\begin{equation}
G_{\mu\nu}\equiv R_{\mu\nu}-\frac{1}{2}R\,g_{\mu\nu}= T^{{\rm
(eff)}}_{\mu\nu} \,,\label{feild:3}
\end{equation}
where the effective stress-energy tensor is defined by $T^{{\rm
(eff)}}_{\mu\nu}=T^{{\rm(m)}}_{\mu\nu}+\tilde{T}^{{\rm(m)}}_{\mu\nu}$.
The latter term is given by
$$\tilde{T}^{{\rm(m)}}_{\mu\nu}=2FT_{\mu\nu}+\left(2\rho
F+f\right)g_{\mu \nu}$$

For the matter content of the wormhole, we consider an anisotropic
fluid source whose stress-energy tensor satisfies the energy
conditions and is given by \cite{Mor88}
\begin{equation}
T_{\mu\nu}=(\rho+p_t)u_\mu \, u_\nu+p_t\,
g_{\mu\nu}+(p_r-p_t)\chi_\mu \chi_\nu \,.\label{stress_tensor}
\end{equation}
Here $\rho$, $p_{t}$ and $p_{r}$ are the energy density, the
perpendicular (to the inhomogeneous direction) pressure, and the
parallel pressure respectively as measured in the fluid element's
rest frame. The vector $u_\mu$ is the fluid four-velocity and
$\chi_\mu$ is a space-like vector orthogonal to $u_\mu$. With these
considerations, the stress-energy tensor takes a diagonal form,
i.e., $T^{\mu}{}_{\nu}={\rm diag}[-\rho(r),p_r(r),p_t(r),p_t(r)]$.
Thus, the gravitational field equations (\ref{feild:3}) are given as
follows
\begin{eqnarray}
\frac{b'}{r^2}=\rho-f \,,  \label{grafield1-t}
       \\
-\frac{b}{r^3}=p_r\left(1+2F\right)+2\rho F+f \,,\label{grafield1-r}
       \\
\frac{b-b'r}{2r^3}=p_t\left(1+2F\right)+2\rho F+f
\,.\label{grafield1-tetta}
\end{eqnarray}
Now we assume that $f(T)=\lambda T$, where $\lambda $ is a constant
\cite{Har11} and with Eq. (\ref{stress_tensor}), $T=-\rho+p_r+2p_t$.
Thus the field equations (\ref{grafield1-t})-(\ref{grafield1-tetta})
yield the following results
\begin{eqnarray}
\rho=\frac{b'}{r^{2}(1+2\lambda)}\,,   \label{grafield1-tt}
 \\
p_r=-\frac{b}{r^3(1+2\lambda)}\,,  \label{grafield1-rr} \\
p_t=\frac{(b-b'r)}{2r^3(1+2\lambda)}\,. \label{grafield1-thetheta}
\end{eqnarray}
These equations describe the matter threading the wormhole, as a
function of the shape function and the coupling parameter $\lambda$.
Note that in the case $\lambda=0$, the general relativistic limit can
be recovered. The system of equations (\ref{grafield1-tt})-
(\ref{grafield1-thetheta}) are three equations with four unknown
functions $\rho(r)$, $p_r(r)$, $p_t(r)$ and $b(r)$. There are
different strategies to solve the field equations. For example, by
specifying an equation of state for the matter field, one can obtain
the shape function and the stress-energy components.

\subsection{Energy Conditions}

As has been mentioned in the introduction, the existence of the
wormhole solution in general relativity relies on the violation of
the null energy condition. The null energy condition holds if
$$T_{\mu\nu}n^\mu n^\nu \ge 0$$ for any null vector field $n^\mu$.
However, if the theory of gravity is chosen to be more complicated
than Einstein gravity, one may circumvent this issue and possess a
throat region which respects energy conditions. Thus, considering a
radial null vector, violation of the Null energy condition, i.e.,
$T_{\mu\nu}^{{\rm (eff)}}\,n^\mu n^\nu < 0$ is given by
\begin{equation}
\rho^{{\rm (eff)}}+p_r^{{\rm (eff)}}=(1+2\lambda)(\rho+p_r)<0.
\end{equation}
On the other hand, with the field equations (\ref{grafield1-tt})-(\ref{grafield1-thetheta})
we deduce that
\begin{equation}
\rho^{{\rm (eff)}}+p_r^{{\rm (eff)}}=\frac{b'r-b}{r^3}\,.
\end{equation}

Using the flaring out condition i.e., $(b'r-b)/b^2<0$, this term is
negative. If we suppose the matter threading the wormhole to satisfy
the energy conditions, imposing the weak energy condition (WEC)
given by $\rho \geq 0$ and $\rho + p_r \geq 0$, we see that the
coupling parameter $\lambda$ is limited to $\lambda \leq
\frac{-1}{2}$.

\subsection{Special Solution:\, $p_{r}=\alpha \rho$}
In this section, we adopt a special equation of state for the matter
field threading the wormhole and obtain the solution of the field
equations (\ref{grafield1-tt})-(\ref{grafield1-thetheta}). An
interesting equation of state is a linear relation between the
radial pressure and the energy density, i.e. $p_{r}=\alpha \rho$,
where $\alpha$ is a constant. Using this equation of state provides
the following shape function
\begin{equation}
b(r)=r_{0}(\frac{r_{0}}{r})^{1/\alpha} . \label{shape}
\end{equation}

\begin{figure}
{\includegraphics[width=3in]{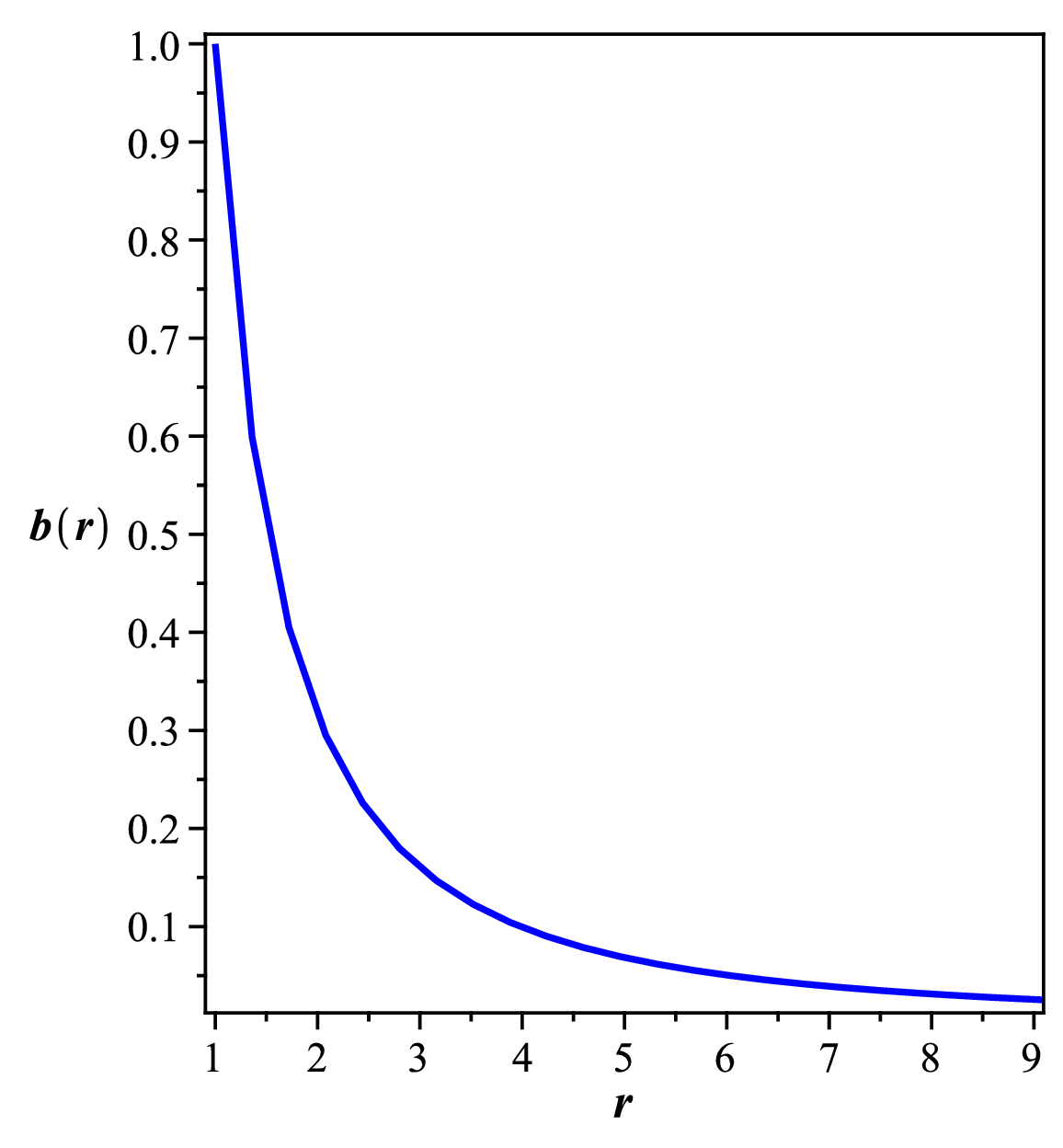}}\hspace{2cm}
\caption{\label{fig:1}The shape function $b(r)$ versus $r$ for
$\alpha=0.6$ and $r_{0}=1$.}
\end{figure}

\begin{figure}
{\includegraphics[width=3in]{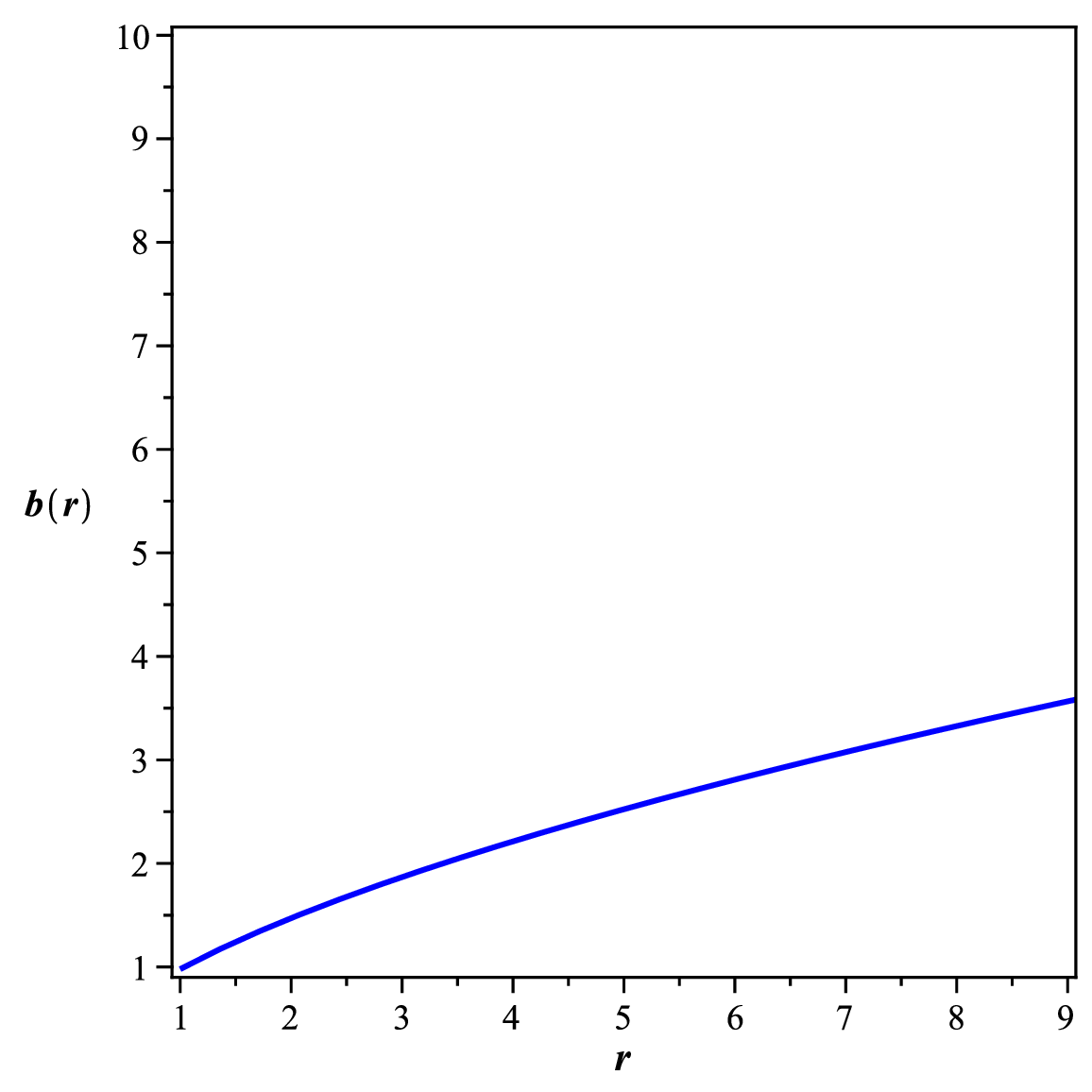}}\hspace{2cm}
\caption{\label{fig:2}The behavior of $b(r)$ versus $r$ for
$\alpha=-1.5$ and $r_{0}=1$.}
\end{figure}

Note that with the condition at the throat $1-\frac{b(r)}{r}\geq0$
(which is equivalent to $r-b(r)\geq0$), the allowed region for the
parameter $\alpha$ is restricted to $\alpha>0$ and $\alpha<-1$. In
Figs. \ref{fig:1} and  \ref{fig:2}, the shape function versus $r$ is
plotted for $\alpha=0.6$ and $\alpha=-1.5$ respectively. As the
figures show, the fundamental wormhole condition, i.e. $b(r)<r$ is
fulfilled.

\begin{figure}
{\includegraphics[width=2.9in]{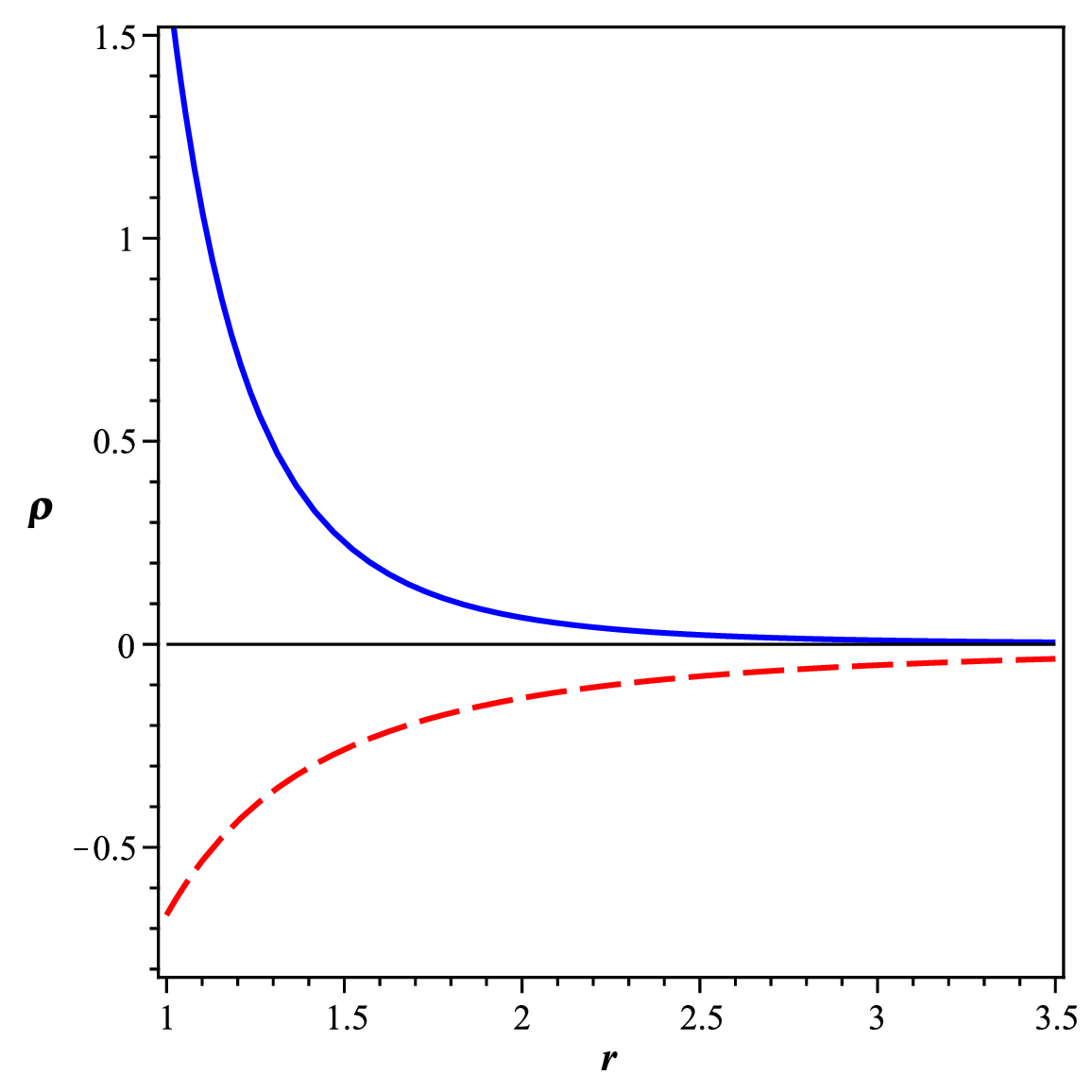}}\hspace{2cm}
\caption{\label{fig:3}The energy density versus $r$ for $\alpha=0.6$
(solid line) and $\alpha=-1.5$ (dashed line) with $\lambda=-1$ and
$r_{0}=1$.}
\end{figure}

Using the shape function (\ref{shape}) with gravitational field
equations (\ref{grafield1-tt})-(\ref{grafield1-thetheta}), the
stress-energy components are given by
\begin{equation}
p_{r}=\alpha \rho=-\frac{Cr^{-(3+1/\alpha)}}{(1+2\lambda)}\,,\label{energy:1}
\end{equation}
and
\begin{equation}
p_{t}=\frac{C(\alpha+1)r^{-(3+1/\alpha)}}{2\alpha(1+2\lambda)}\,,\label{energy:2}
\end{equation}
where $C=r_{0}^{1+1/\alpha}$. Fig. \ref{fig:3} shows the energy
density versus $r$ for $\alpha=0.6$ and $\alpha=-1.5$. Obviously,
choosing a negative value for $\alpha$ in the allowed region, leads
to a negative energy density for the matter threading the wormhole.

\begin{figure}
{\includegraphics[width=3in]{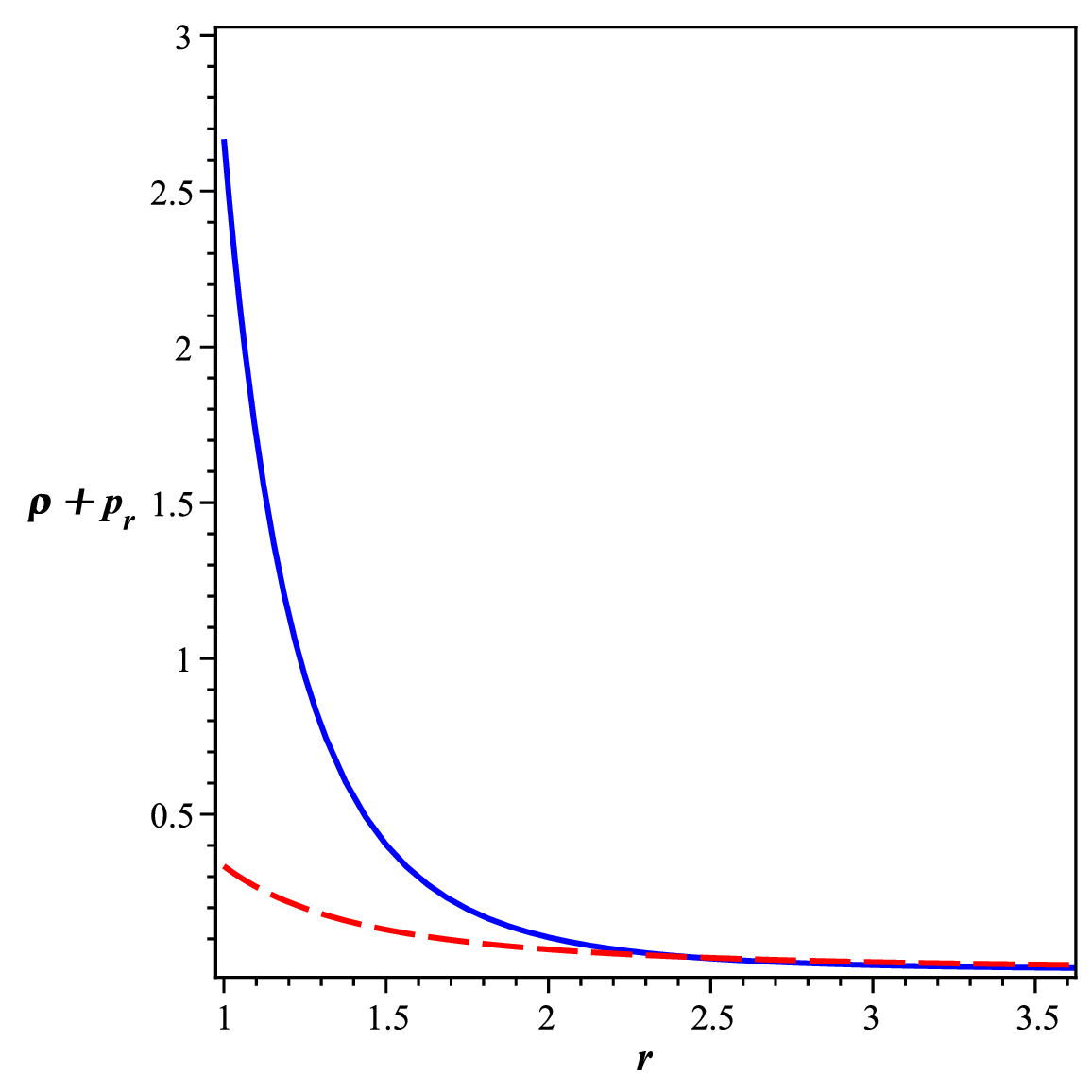}}\hspace{2cm}
\caption{\label{fig:4}The null energy condition is respected for
$\alpha=0.6$ (solid line) and $\alpha=-1.5$ (dashed line) with
$\lambda=-1$ and $r_{0}=1$.}
\end{figure}

Now, using expression (\ref{energy:1}), the region satisfying the
NEC condition at the throat is depicted in Fig. \ref{fig:4} for
$\alpha=0.6$ and $\alpha=-1.5$ respectively. As the figures show,
the stress-energy tensor satisfies the null energy condition. As a
result, for a positive value of $\alpha$, the weak energy condition
is satisfied. On the other hand, by choosing a negative value for
$\alpha$ in the allowed region, only the null energy condition is
respected.

\section{Conclusion}
The existence of the traversable wormholes as solutions to the
Einstein field equations relies on the presence of some form of
exotic matter which violate the null energy condition. However, in
the framework of a modified theory of gravity, the situation may be
completely different. In this paper, we have investigated wormhole
solutions in $f(R,T)$ modified gravity where $R$ is the curvature
scalar and $T$ is the trace of the stress-energy tensor. We have
shown that in the context of $f(R,T)$ gravity, traversable wormhole
solutions can be obtained, without the need to introducing any form
of exotic matter. Violation of the energy conditions, which is
essential for the existence of the wormhole solutions \cite{Mor88},
is realized via the presence of an effective stress-energy tensor
generated by the additional curvature and matter terms.

To find the wormhole solutions in $f(R,T)$ scenario, we have assumed
that $f\left(R,T\right)=R+2f(T)$, where $f(T)$ is an arbitrary
function of the trace of the stress-energy tensor and we have
derived the gravitational field equations. Then we have specified an
equation of state for the matter threading the wormhole and by
imposing the flaring out condition at the throat, we have obtained
the shape function. We have shown that the stress-energy tensor of
the matter threading the wormhole satisfies the null energy
condition in some subspaces of the model parameter space. However,
one can explore the wormhole solutions in $f(R,T)$ modified gravity
in more complicated situation than here. For example, in a recent
work \cite{Azi12}, we investigated the wormhole solutions in the
case that the gravity sector is also modified and we studied the
energy conditions in this case. Furthermore one can consider an
explicit coupling between $T$ and $R$ and explore the wormhole
geometries.

\section*{Acknowledgements}
The author would like to acknowledge Prof. Kourosh Nozari for
invaluable remarks. We also thank Prof. F. S. Lobo for useful
discussions.

\end{document}